\documentclass{INTERSPEECH2023}

\interspeechcameraready 

\usepackage{amsmath,graphicx}
\usepackage{subfig}

\renewcommand{\vec}[1]{\mbox{\boldmath${#1}$}}

\title{APPLICATION OF KNOWLEDGE DISTILLATION TO MULTI-TASK SPEECH REPRESENTATION LEARNING}

\name{$^{*}$Mine Kerpicci$^{1}$ \thanks{$^{\star}$ Work performed as an intern at Qualcomm Technologies} \qquad Van Nguyen$^{2}$ \qquad Shuhua Zhang$^{2}$\qquad Erik Visser$^{2}$}

\address{$^{1}$ Georgia Institue of Technology, Atlanta, Georgia \\
	$^{2}$ Qualcomm Technologies, Inc., San Diego, California }

\begin{document}

\maketitle
 
\begin{abstract}
Model architectures such as wav2vec 2.0 and HuBERT have been proposed to learn speech representations from audio waveforms in a self-supervised manner. When they are combined with downstream tasks such as keyword spotting and speaker verification, they provide state-of-the-art performance. However, these models use a large number of parameters, the smallest version of which has 95 million parameters. This constitutes a challenge for edge AI device deployments. In this paper, we investigate the application of knowledge distillation to speech representation learning (SRL) models followed by joint fine-tuning with multiple downstream voice-activated tasks. In our experiments on two such tasks, our approach results in nearly 75\% reduction in model size while suffering only 0.1\% accuracy and 0.9\% equal error rate degradation compared to the full-size model. In addition, we show that fine-tuning the SRL models results in a significant performance boost compared to using frozen SRL models.
\end{abstract}

\section{Introduction}
\label{sec:intro}

Speech representation learning (SRL) has been extensively studied in the literature and shown potential for various speech recognition tasks \cite{srl_1, srl_2, srl_4}. Deep neural network architectures \cite{deep, deep2} are the method of choice for SRL. Since sufficient amounts of labeled data are not  typically available for supervised learning in real scenarios, this problem is circumvented by using self-supervised learning on unlabeled data \cite{wav2vec, hubert, tera, selfsupervised_1, selfsupervised_2}. In \cite{wav2vec, hubert}, convolutional networks are combined with transformer encoder-based models to learn speech representations in a self-supervised manner. \cite{emotionrecognition, emotionrecognition_example, speaker_recognition} propose to combine such a model with a downstream network to address classification problems such as emotion recognition or speaker recognition. The resulting models provide competitive performance on these tasks but require training multiple large networks each for a different single task. To enable models for multi-tasking, \cite{multitask} proposes to learn speech representations from raw audio with wav2vec 2.0 \cite{wav2vec} and then fine-tune the model for multiple tasks simultaneously by sampling training instances from the combined task-specific datasets. This method provides even better performance than training on single tasks separately.

However, the memory footprint of the resulting models is large with the  number of parameters ranging from 95 million to 1 billion, which makes them difficult to deploy on edge AI devices. This problem is studied with parameter sharing in \cite{review_w2v2}. Knowledge distillation is another method of choice to address model complexity reduction in various fields ranging from natural language processing to speech recognition \cite{kd_survey, kd_app2, kd_app1}. A common approach in the literature is to reduce model size by maintaining its performance via teacher-student network distillation architectures. \cite{shrinking,review_lighthubert} apply knowledge distillation to large SRL methods to construct efficient speech recognition networks. The conventional approach is transferring knowledge from the final layer of the teacher network to the final prediction of the student network. However, the middle layers of these SRL models also contain valuable speech representation information \cite{layer_analysis}. In \cite{distilhubert}, layer-wise distillation to the HuBERT \cite{hubert} model is applied where knowledge is transferred from middle layers of the large network to construct a smaller model.

\begin{figure*}[t]
	\centering
	\subfloat[\label{fig:distillation_flow}]{%
		\includegraphics[width=0.5\textwidth]{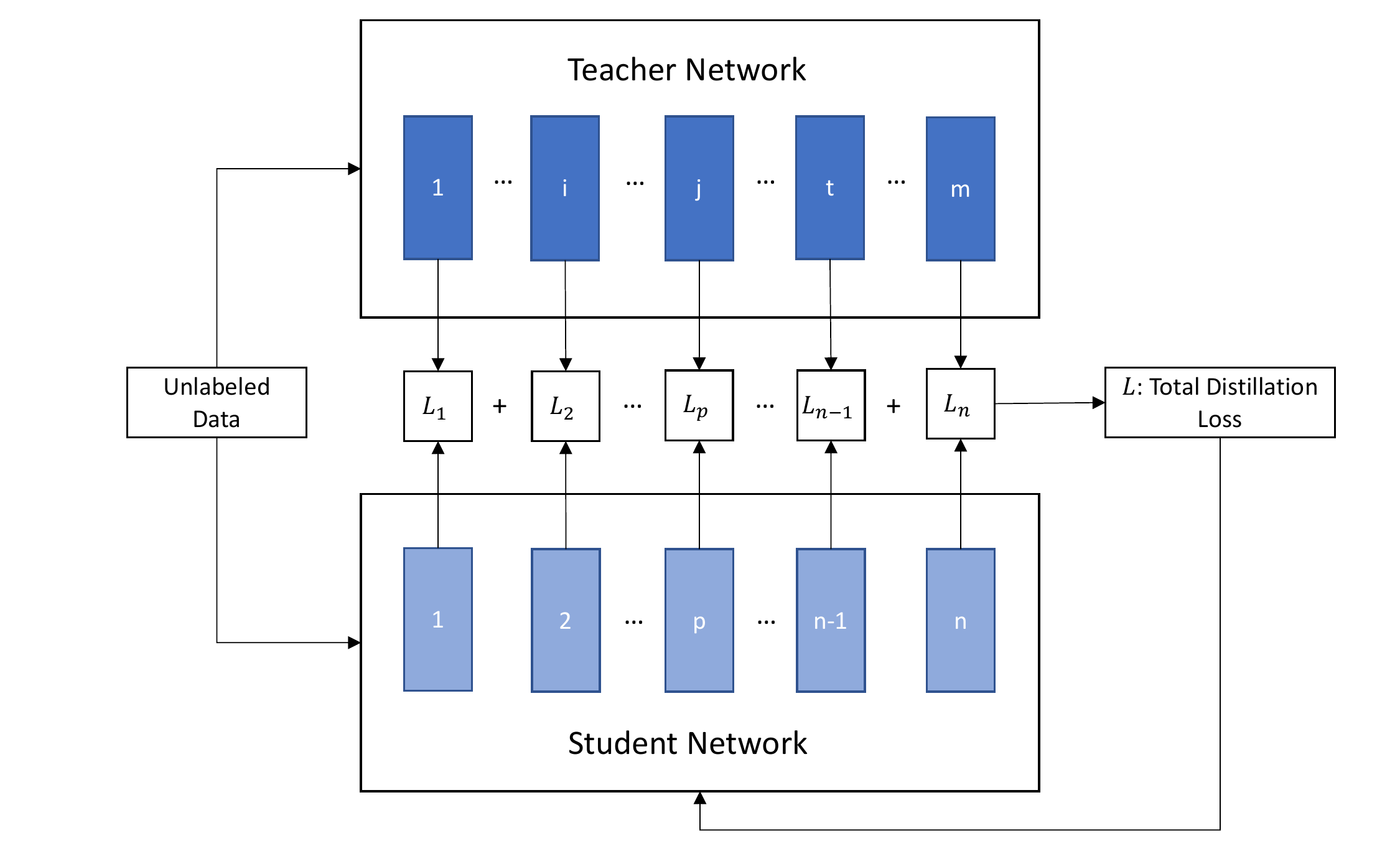}
	}
	\subfloat[\label{fig:finetuning_flow}]{%
		\includegraphics[width=0.5\textwidth]{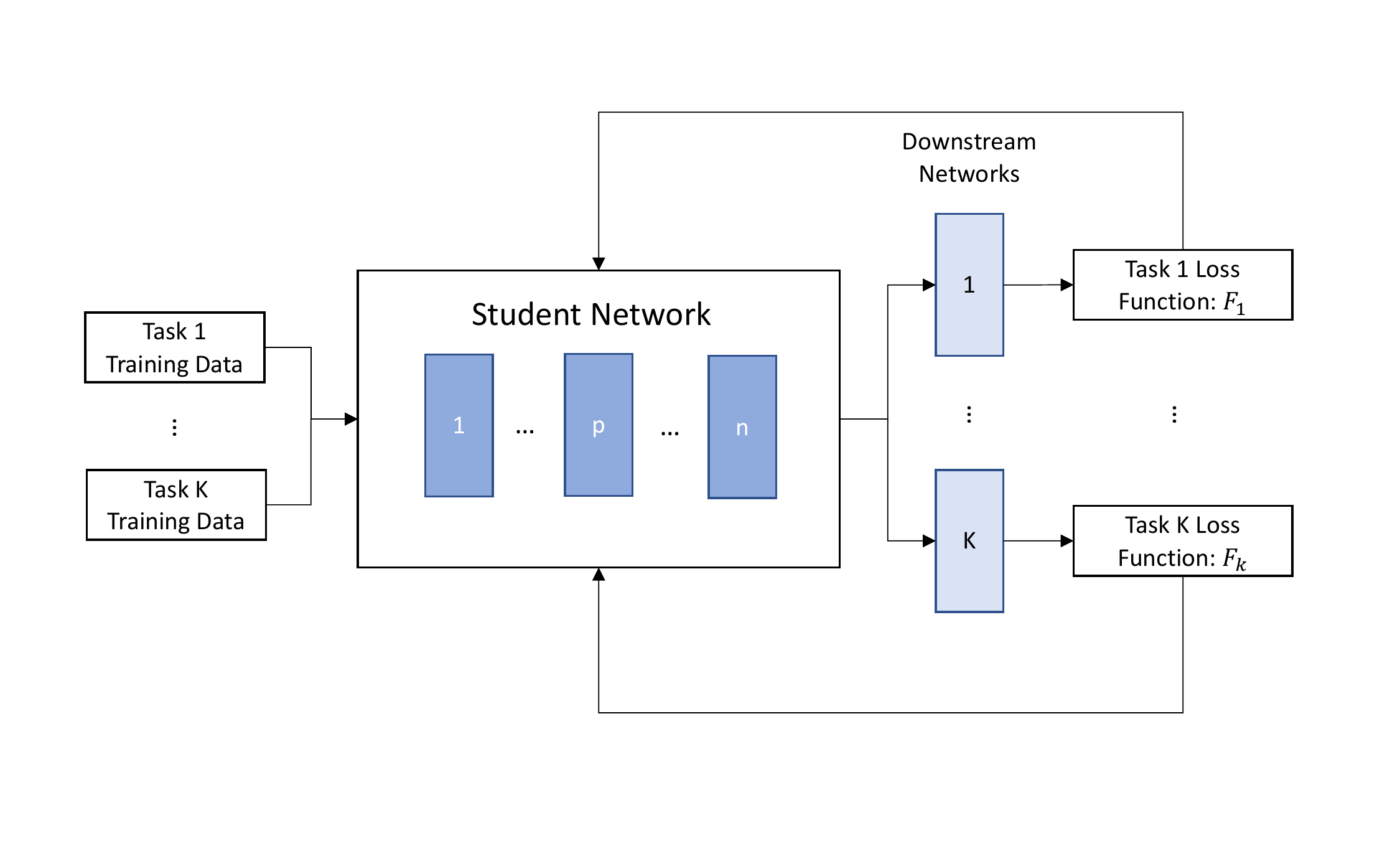}
	} 
	\caption{
		Flowchart of 
		(a) the knowledge distillation process where the student network is trained via hidden layers of the pre-trained teacher network (distillation can be performed among any desired layers with a suitable loss function choice),
		(b) the fine-tuning step where the distilled network is trained with a multi-task training scheme.} 
	\label{fig:flowchart}
\end{figure*}

In our work, we discuss generalization of knowledge distillation to use a large network in training of another network with reduced model size, and we use a special knowledge distillation approach as presented in \cite{distilhubert} in our experiments.
As opposed to this approach, we combine the student model with linear downstream heads and fine-tune all network parameters via a multi-task training scheme as in \cite{multitask} instead of freezing the SRL model as done in \cite{distilhubert}. In our experiments, we address keyword spotting and speaker verification problems in single task and multi-task frameworks. Through our results, we show that the constructed student networks perform as well as the teacher models even though the former model size is 28\% of the latter. 
Moreover, we perform this in two settings where we construct distilled SRL modules with wav2vec 2.0 and HuBERT separately for a comprehensive performance comparison.

To the best of our knowledge, we propose for the first time the combination of knowledge distillation application to SRL and joint fine-tuning of the complete model (SRL module and downstream heads) for multiple downstream voice-activated tasks.
This is in contrast to common speech recognition multi-task training approaches where an SRL module is frozen, functioning as a feature extractor, and only the downstream speech recognition tasks are trained \cite{distilhubert, superb}.
On the contrary, we show that fine-tuning both distilled SRL module and downstream heads for multiple voice-activated tasks simultaneously achieves significantly higher performance on all such tasks compared to fine-tuning only the downstream heads with frozen SRL module.

The main contributions of this work are as follows:

\begin{itemize}
	\item We successfully combine knowledge distillation and multi-task training to construct a single network (with about 75\% reduced size) that can be efficiently used in devices and embedded systems for multiple tasks.
	\item We show that training the complete network end-to-end (both SRL module and downstream heads) on multiple tasks achieves significantly higher performance than training on the same tasks with the frozen SRL module.
	\item We discuss generalization of the used methods (knowledge distillation and multi-task training) and show that they can be efficiently applied to different networks where we achieve competitive results with both distilled wav2vec 2.0 and distilled HuBERT models on keyword spotting (KWS) and speaker verification (SV) voice-activated tasks.
\end{itemize}

The paper is organized as follows. Section \ref{sec:method} describes our approach in detail, Section \ref{sec:exp} presents experiments conducted to validate our approach and we conclude with Section \ref{sec:conclusion}.

\section{METHOD}
\label{sec:method}

We apply the student-teacher approach \cite{kd_survey} to construct our SRL module.
We use wav2vec 2.0 and HuBERT models as our teacher network as discussed in Section \ref{sec:srl} and apply distillation method explained in Section \ref{sec:kd}. Then, we fine-tune the student model and downstream network with the multi-task training approach in Section \ref{sec:multitask}. We provide the flowchart of the complete process in Figure \ref{fig:flowchart}. Note that the proposed end-to-end training scheme can be applied to various teacher-student models for several tasks. 
Here, in particular, we investigate its application to SRL models and present our experimental analysis on multiple voice-activated tasks.

\subsection{Speech Representation Learning}
\label{sec:srl}

HuBERT \cite{hubert} and wav2vec 2.0 \cite{wav2vec} methods are capable of learning powerful speech representations from audio waveforms in a self-supervised manner. 
They share the same architecture consisting of CNN and transformer encoder layers. Their BASE model corresponding to the smallest version contains 95 million parameters while LARGE model contains 317 million parameters. HuBERT has an additional X-LARGE model with 1 billion parameters. While sharing a similar structure, differences result from their pre-training schemes. The wav2vec 2.0 model has a single step of training with a quantization process to learn the targets. 
HuBERT, however, first constructs the pseudo-targets in a separate clustering step and then learns these targets in the second step during training. 
As a result, each of their hidden layers learns different properties of speech representation and transfers various abilities when they are used for distillation.

In our experiments, we investigate both wav2vec 2.0 and HuBERT as our teacher network for knowledge distillation to construct separate student networks. Then, we evaluate these in two settings to compare the performance of final student models and to show the applicability of the discussed processes.

\begin{figure*}[t]
	\centering
	\subfloat[\label{fig:distillation_diag}]{%
		\includegraphics[width=0.5\textwidth]{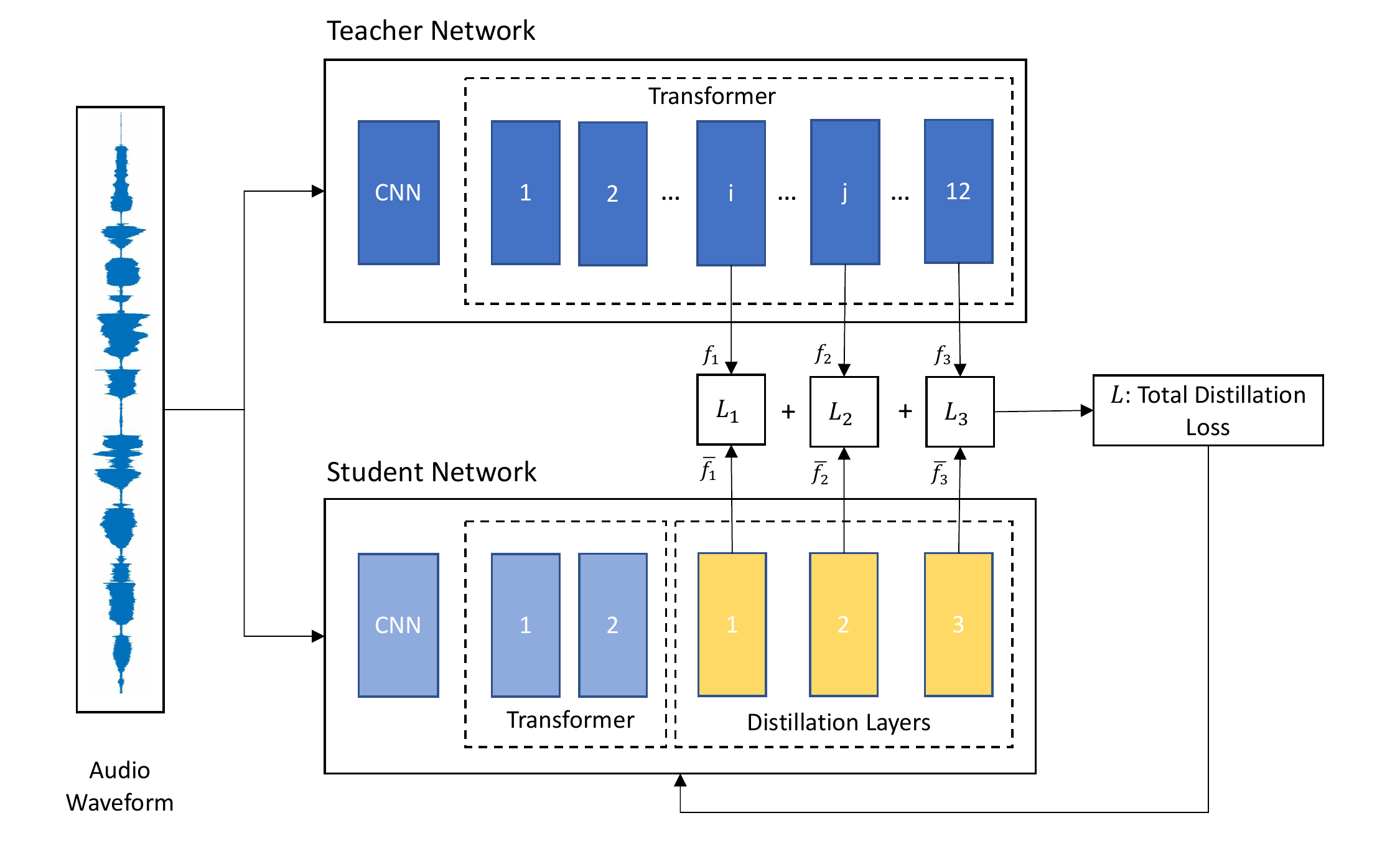}
	}
	\subfloat[\label{fig:finetuning_diag}]{%
		\includegraphics[width=0.5\textwidth]{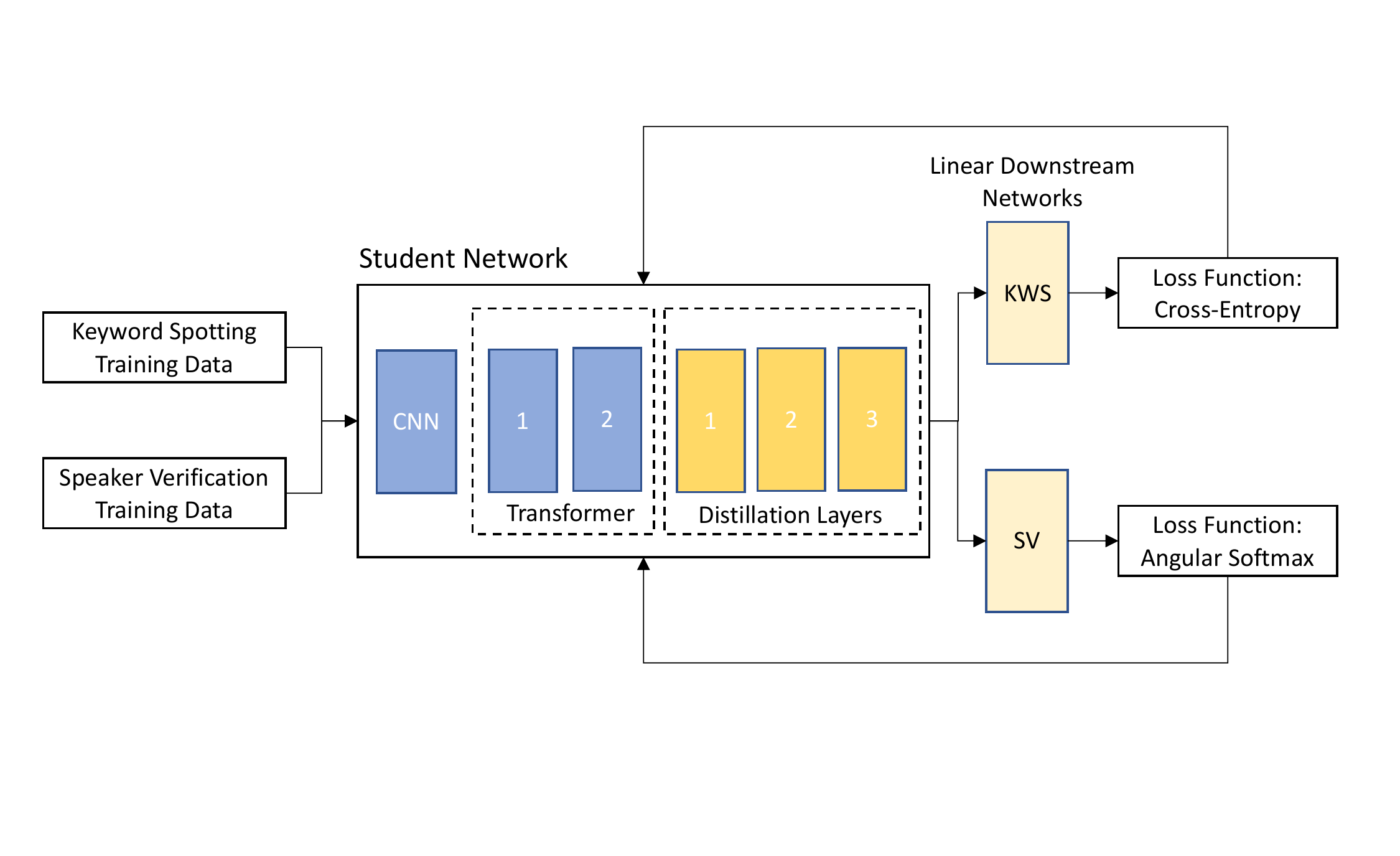}
	} 
	\caption{Flowchart of 
		(a) the distillation process where the student model with CNN and Transformer layers is trained via teacher network layers,
		(b) the fine-tuning step where the complete distilled network is trained with a multi-task training scheme for keyword spotting and speaker verification tasks.} 
	\label{fig:diagram}
\end{figure*}

\subsection{Knowledge Distillation}
\label{sec:kd}

Deep architectures achieve state-of-the-art performance in many tasks due to their high learning capacities. However, their deployment in mobile devices and embedded systems is a challenging problem due to cost and memory limitations. In the literature, this problem is addressed with the idea of transferring knowledge from such models to smaller networks to reduce network size while keeping the performance similar.

Knowledge distillation is a method that transfers knowledge from layers of a large pre-trained network (teacher) to a simpler or smaller model (student) as demonstrated in Figure \ref{fig:distillation_flow}. As in this generalized representation, the transfer can be performed among any layers of the teacher and the student networks where the teacher guides the student through a defined loss function during pre-training \cite{kd_survey}.
Several studies have suggested that not only the final layer but also the middle layers of SRL modules contain valuable information \cite{layer_analysis, layer_info_2}. Hence, it is proposed to transfer knowledge from middle layers of teacher as well as its final layer to learn the hidden representations \cite{distilhubert}. We use this approach in our pre-training as a special case of the process in Figure \ref{fig:distillation_flow}. 
In this method, knowledge is transferred between only some of the student and teacher layers via feature vectors as in the diagram in Figure \ref{fig:distillation_diag}. 
The student parameters are initialized with their corresponding layers in the teacher network and trained with the loss function
\begin{align*}
	L_{p} = \sum_{i=1}^{T} \bigg( \frac{1}{K}|| \vec{f}_{p, i} -  \bar{\vec{f}}_{p, i} ||_{1} - \log \sigma \Big( \cos (\vec{f}_{p, i} , \bar{\vec{f}}_{p, i})  \Big) \bigg).
\end{align*}
This function is separately calculated for all hidden layers where distillation is performed. Here, $T$ is the number of time steps and $K$ is the dimension of feature vectors where $\vec{f}_{p}$ and $\bar{\vec{f}}_{p}$ are the feature vectors of the $p^{\text{th}}$ distillation layer of the teacher and student models, respectively.
Moreover, $\sigma$ is sigmoid function and $\cos$ is cosine similarity.
The aim of this loss function is to minimize $\text{L}_{1}$ distance while maximizing the cosine similarity between the feature vectors.
Note that this method can be applied to the networks in various forms by changing the loss functions accordingly.

\subsection{Multi-task Training}
\label{sec:multitask}

We combine the pre-trained student network with downstream networks to perform multi-task training. For this, we use separate downstream heads for KWS and SV tasks. Note that the choice of downstream heads can vary based on the applications. In our study, we use a linear layer that maps the output of student network to task specific vectors as in Figure \ref{fig:finetuning_diag} to keep the network size small.

In our work, we apply a special multi-task training method where we fine-tune all network parameters (both distilled student network and downstream heads) for all tasks simultaneously \cite{multitask}. In each iteration, we first train all parameters based on the cross-entropy loss function with the KWS dataset and then use the angular softmax loss with the SV dataset. We alternate between the loss functions and batches of the tasks at each iteration. This alteration between both tasks continues until the defined maximum number of iterations is reached. 

Note that the concept of multi-task learning in the literature is commonly considered as fine-tuning multiple downstream heads on their own datasets after freezing the SRL module that they are sharing \cite{distilhubert, superb}. However, in this study, we discuss the multi-task training scheme where the SRL module is also trained with the datasets of all downstream tasks one-by-one during fine-tuning. This training scheme allows the SRL module to adapt its learning capability to the downstream tasks more while increasing its performance on all trained tasks.
In our experiments, we provide the performance results of both training approaches and show that applying this multi-task training scheme without freezing the SRL module achieve significant performance gain in all investigated setups and tasks.

\section{EXPERIMENTS}
\label{sec:exp}

\subsection{Experimental Setup}
\label{sec:setup}

In our experiments, we evaluate the performance of the student networks after distillation on single task and multi-task frameworks. 
We perform our experiments in two settings where we use different models as teacher networks.  
For knowledge distillation, we use BASE models of wav2vec 2.0 \cite{wav2vec} and HuBERT \cite{hubert}, where they have the same architecture of CNN and 12 transformer layers resulting in 95 million parameters. Note that they are both pre-trained with 960-hour LibriSpeech data \cite{librispeech} to learn speech representations in their own training settings. 

The student networks are smaller and simplified versions of the teacher architectures. 
They consist of CNN and 2 transformer layers as in Figure \ref{fig:diagram} where their parameters are initialized with the parameters of CNN and $1^{\text{st}}$ and $2^{\text{nd}}$ transformer layers of the teacher network.
Then, we perform the knowledge distillation from the layers $4$, $8$ and $12$ with the same 960-hour LibriSpeech data. 
The resulting student models with distillation and downstream layers contain 27 million parameters. We refer to these distilled student networks with downstream heads as ``DistilledWav2vec" and ``DistilledHuBERT". 
Note that we keep the layers after distillation, and we fine-tune all network parameters including CNN and transformer layers on downstream tasks. 
For comparison, we also include the performance of the models when all layers are frozen except the downstream heads during fine-tuning step which is called as partial fine-tuning. We refer to these versions as ``DistilledWav2vec (frozen)" and ``DistilledHuBERT (frozen)". The properties of the constructed models regarding the number of parameters and their fine-tuned parts are provided in Table \ref{table:summary} with their model names as referred in the experimental results.

\begin{table}[t]
	\centering
	\caption{The models constructed with the proposed approach and evaluated in the experiments where their number of parameters and fine-tuned network parts are presented.} 
	\label{table:summary}
	\resizebox{1\columnwidth}{.17\columnwidth}
	{\begin{tabular}{l|l|l}
			\hline
			Model              &
			\multicolumn{1}{c|}{\begin{tabular}[c]{@{}c@{}}   {$\#$} \\ Parameters \end{tabular}}  & \multicolumn{1}{c}{\begin{tabular}[c]{@{}c@{}}Fine-tuned Parameters \end{tabular}}  \\ \hline \hline
			wav2vec 2.0  & 96 M   &  SRL Module + Downstream Heads            \\
			wav2vec 2.0 (frozen)   & 96 M           & Downstream Heads     \\ \hline
			DistilledWav2vec   & 27 M   &   SRL Module + Downstream Heads     \\
			DistilledWav2vec  (frozen) & 27 M           & Downstream Heads                \\
			DistilledHuBERT    & 27 M   &  SRL Module + Downstream Heads       \\
			DistilledHuBERT   (frozen) & 27 M            & Downstream Heads                  \\ \hline
	\end{tabular}}
\end{table}

We perform our experiments in both single task and multi-task frameworks for a comprehensive analysis. In these experiments, we use keyword spotting and speaker verification tasks to train the networks for various voice-activated tasks.

For keyword spotting (KWS), we use the Google Speech Command (GSC) V1 dataset \cite{googlespeechcommands} with 22236 train and 3081 test utterances of 12 classes and provide accuracy results on the test set. Also, we use VoxCeleb-1 dataset \cite{voxceleb} with 143642 train and 4874 test utterances for speaker verification (SV) with Equal Error Rate (EER) evaluation metric. In the experiments, we use Adam optimizer with a learning rate $10^{-4}$. Note that we also provide the performance of teacher wav2vec 2.0 model as a reference since our aim is to achieve the performance of such a powerful and large network with much smaller networks.

\subsection{Single Task Results}
\label{sec:results1}

In this section, we provide our experimental results regarding the single task setting. We construct two different student networks as DistilledWav2vec and DistilledHuBERT from different teachers. We train them separately for single tasks of KWS and SV. Note that this is different from the multi-task training scheme described in Section \ref{sec:multitask}. Here, the networks are trained for only one task at each time. 

We provide the evaluation results on test sets in Table \ref{table:singletask}. 
As seen from the accuracy and EER values, the student models perform well on both KWS and SV tasks. 
Particularly, we observe only 0.1\% accuracy degradation on the KWS task with DistilledWav2vec compared to the teacher wav2vec 2.0. Moreover, we obtain only 0.9\% EER degradation on SV with DistilledHuBERT even though the student networks' size is only 28\% of the large teacher network.

In this set of experiments on single task, we show that the knowledge distillation method can be efficiently applied to various networks to reduce the model size while maintaining the performance. However, these models are trained to work on only one task. 
In the next section, we aim to achieve similar competitive results on multiple tasks with a single network.

\begin{table}[t]
	\centering
	\caption{The results of single task experiments performed on keyword spotting and speaker verification tasks seperately.} 	
	\label{table:singletask}
	\resizebox{1\columnwidth}{.19\columnwidth}
	{\begin{tabular}{l|l|l|l|l}
			\hline
			Model              &
			\multicolumn{1}{c|}{\begin{tabular}[c]{@{}c@{}}   {$\#$} \\ Parameters \end{tabular}} & \multicolumn{1}{c|}{\begin{tabular}[c]{@{}c@{}}Training \\ Task\end{tabular}} & \multicolumn{1}{c|}{\begin{tabular}[c]{@{}c@{}}KWS \\ Accuracy (\%)\end{tabular}} & \multicolumn{1}{c}{\begin{tabular}[c]{@{}c@{}}SV\\ EER (\%)\end{tabular}} \\ \hline \hline
			wav2vec 2.0 \cite{multitask}  & 96 M &
			KWS           & 98.10               & -             \\
			wav2vec 2.0 \cite{multitask}  & 96 M & SV            & -                   & 3.35          \\ \hline
			DistilledWav2vec   & 27 M &  KWS           & \textbf{98.01 }               &         -      \\
			DistilledWav2vec   & 27 M & SV            & -                   & 4.48         \\
			DistilledHuBERT    & 27 M & KWS           & 97.82              &         -      \\
			DistilledHuBERT    & 27 M & SV            & -                   & \textbf{4.26 }          \\ \hline
	\end{tabular}}
\end{table}

\subsection{Multi-Task Results}
\label{sec:results2}

We perform multi-task training on the student networks DistilledWav2vec and DistilledHuBERT. In this case, we train the networks with KWS and SV tasks simultaneously where the resulting networks can be used in devices with multi-task applications efficiently. 

We obtain the evaluation results as in Table \ref{table:multitask}. In this setting, training the complete network (both SRL module and dowsntream heads) instead of freezing the SRL module provides 8.41\% and 6.63\% accuracy gains on KWS, respectively for DistilledWav2vec and DistilledHuBERT models. Similarly on the SV task, these models obtain 10.42\% and 8.40\% EER improvements due to the discussed multi-task training scheme.
Based on these accuracy and EER values, we observe that entire fine-tuned models perform better than partial fine-tuned versions in all cases. Hence, we conclude that training not only downstream heads but also SRL modules with the multi-task training scheme significantly increases the overall performance.

Moreover, we observe that both DistilledHuBERT and DistilledWav2vec models that are trained in a multi-task framework provide similar results with their single task comparisons. Particularly, when DistilledHuBERT is trained solely on the KWS task, it achieves 97.82 {\%} accuracy and when it is trained solely on SV, it has 4.26 {\%} EER as in Table \ref{table:singletask}. On the other hand, when we fine-tune DistilledHuBERT on both KWS and SV at the same time, we achieve 97.90 {\%} accuracy on KWS and 4.27 {\%} EER on SV as in Table \ref{table:multitask}. This shows that, instead of using a different network for each task, we can achieve similar competitive performance results on multiple tasks with a single network by applying fine-tuning to both SRL module and downstream heads in a multi-task training scheme. 
Hence, the resulting network can be efficiently deployed and used for multiple tasks by achieving high performance with low memory and computational costs.

Finally, we conclude that both DistilledHuBERT and DistilledWav2vec student networks after entire fine-tuning in multi-task training setting reach competitive performances. Particularly for KWS task, the performance gap between teacher and student networks is small. This means that we can match the large network performance with a much smaller model
even in a multi-task application framework.

\begin{table}[t]
	\centering
	\caption{The results of multi-task experiments performed simultaneously for keyword spotting and speaker verification.} 
	\label{table:multitask}
	\resizebox{1\columnwidth}{.19\columnwidth}
	{ \begin{tabular}{l|l|l|l}
			\hline
			Model                  
			&
			\multicolumn{1}{c|}{\begin{tabular}[c]{@{}c@{}}   {$\#$} \\ Parameters \end{tabular}} &
			\multicolumn{1}{c|}{\begin{tabular}[c]{@{}c@{}}   KWS \\ Accuracy (\%) \end{tabular}} & 
			\multicolumn{1}{c}{\begin{tabular}[c]{@{}c@{}}   SV \\ EER (\%) \end{tabular}} 
			\\ \hline \hline
			wav2vec 2.0 \cite{multitask}         & 96 M & 98.19               & 3.15          \\
			wav2vec 2.0 (frozen) \cite{multitask} & 96 M & 90.33               & 20.5          \\ \hline
			DistilledWav2vec            & 27 M & 97.70               & 4.59          \\
			DistilledWav2vec (frozen)   & 27 M & 89.29               & 15.01         \\
			DistilledHuBERT             & 27 M & \textbf{97.90}      & \textbf{4.27} \\
			DistilledHuBERT (frozen)    & 27 M & 91.27               & 12.67         \\ \hline
	\end{tabular}}
\end{table}

\section{CONCLUSION}
\label{sec:conclusion}

In this paper, we apply a knowledge distillation approach to different self-supervised networks to obtain high speech representation learning performance with reduced network size and illustrate its generalization to different settings. We perform single task and multi-task trainings for various speech recognition applications and provide a comprehensive performance comparisons. 
Particularly, we propose to combine knowledge distillation and multi-task training scheme where the complete network of SRL module and downstream heads is fine-tuned with all task datasets simultaneously to obtain a single network that can be efficiently used for multiple tasks.
In our experiments, we show that the performance of large models such as wav2vec 2.0 and HuBERT can be achieved with substantially smaller networks (28\% of the original model sizes) by using the training approach presented.
Moreover, we show that the performance of single task networks on their particular tasks can be achieved with a single network fine-tuned on multiple tasks. Hence, these multi-task models do not only provide high performance, but also reduce memory and computation costs by allowing a single network to be shared among multiple tasks.
The resulting networks are thus potential candidates for deployment on edge AI devices while offering competitive performance.

\bibliographystyle{IEEEtran}
\bibliography{mybibfile}

\end{document}